\documentclass{bmvc2k}

\title{Adaptive Lighting for Data-Driven Non-Line-of-Sight 3D Localization and Object Identification}

\addauthor{Sreenithy Chandran}{schand56@asu.edu}{1}
\addauthor{Suren Jayasuriya}{sjayasur@asu.edu}{1}

\addinstitution{
 Arizona State University\\
 Tempe, USA
}

\runninghead{Chandran \& Jayasuriya}{Adaptive Lighting for NLOS Imaging}
\usepackage{times}
\usepackage{epsfig}
\usepackage{graphicx}
\usepackage{amsmath}
\usepackage{amssymb}
\usepackage{makecell}
\usepackage[norelsize,ruled]{algorithm2e}
\usepackage{color,soul}
\usepackage{array}
\newcolumntype{P}[1]{>{\centering\arraybackslash}p{#1}}
\newcolumntype{M}[1]{>{\centering\arraybackslash}m{#1}}

\DeclareMathOperator*{\argmax}{argmax}

\begin{document}

\maketitle

\begin{abstract}
  Non-line-of-sight (NLOS) imaging of objects not visible to either the camera or illumination source is a challenging task with vital applications including surveillance and robotics.  Recent NLOS reconstruction advances have been achieved using time-resolved measurements which requires expensive and specialized detectors and laser sources. In contrast, we propose a data-driven approach for NLOS 3D localization and object identification requiring only a conventional camera and projector. To generalize to complex line-of-sight (LOS) scenes with non-planar surfaces and occlusions, we introduce an adaptive lighting algorithm. This algorithm, based on radiosity, identifies and illuminates scene patches in the LOS which most contribute to the NLOS light paths, and can factor in system power constraints. We achieve an average identification of 87.1\% object identification for four classes of objects, and average localization of the NLOS object's centroid with a mean-squared error (MSE) of 1.97 cm in the occluded region for real data taken from a hardware prototype. These results demonstrate the advantage of combining the physics of light transport with active illumination for data-driven NLOS imaging.  
\end{abstract}

% !TEX root = main.tex

\section{Introduction}

Non-line-of-sight (NLOS) imaging is an emerging field of research with applications for autonomous vehicle collision avoidance, search and rescue operations, and industrial safety and inspection. This problem is challenging because NLOS objects are outside of the line-of-sight (LOS) of both the camera and the illumination source(s). Recent success in NLOS reconstruction utilizes time-of-flight to help separate NLOS from LOS path contributions~\cite{velten2012recovering,buttafava2015non,o2018confocal}. Using time-resolved measurements, backpropagation~\cite{arellano2017fast} and/or optimization~\cite{heide2014diffuse} can reconstruct the NLOS scene, typically under the assumption of a flat LOS wall to avoid LOS indirect light. These time-resolved detectors, while achieving superior results, are costly, consume higher power than conventional CMOS image sensors, and require specialized aligned optics and lasers. To alleviate these issues, research has also focused on NLOS imaging using conventional cameras~\cite{klein2016tracking,bouman2017turning, torralba2014accidental,thrampoulidis2018exploiting, tancik2018data}.

\begin{figure}
\setlength\abovecaptionskip{-0.7\baselineskip}
 \begin{center}
\includegraphics[width=1\columnwidth]{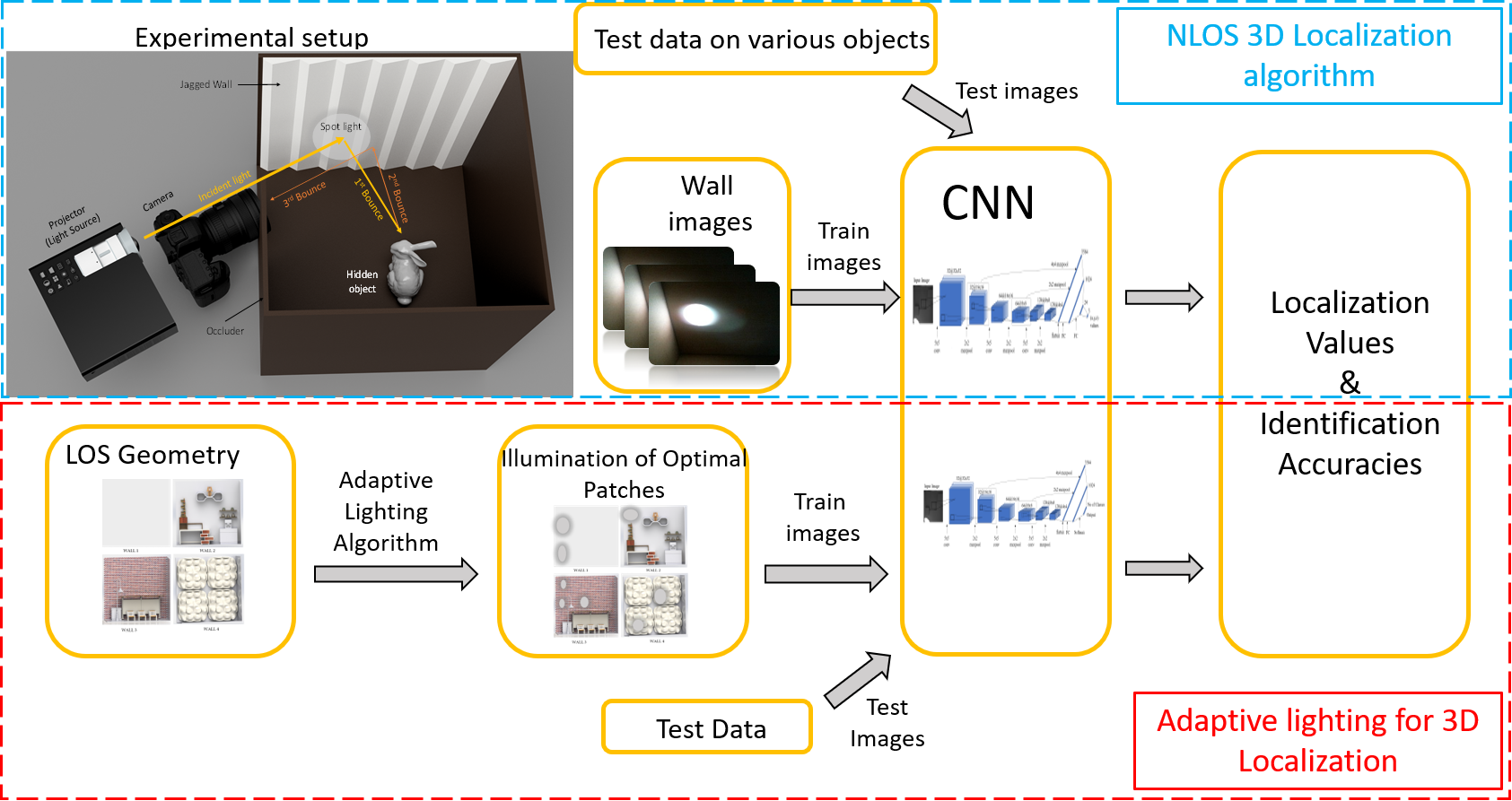}
 \end{center}

  \caption{Our research uses deep learning to perform 3D localization of an object in the non-line-of-sight (NLOS) of a camera and projector. Our complete system pipeline consists of a CNN trained on diffuse wall images, and the adaptive lighting procedure to improve performance. We use two different network architectures for localization and identification. }
\label{fig:problem}
\end{figure}

We propose a data-driven approach to the problem of NLOS 3D localization as well as object identification that leverages modern existing convolutional neural network architectures. This approach requires only a conventional 2D camera and a projector, and no optical alignment or projector-camera synchronization. To handle LOS scenes with non-planar geometries and occlusions, we develop an adaptive lighting algorithm based on radiosity that improves our learning performance. To the best of our knowledge, we are the first paper to perform NLOS imaging on non-planar, complex LOS scenes. 

In particular, our specific contributions are (1) solving NLOS 3D localization as well as NLOS object identification using convolutional neural networks with a conventional camera and projector, (2) an adaptive lighting algorithm for identifying optimal LOS patches to maximize NLOS radiosity captured, and distribute illumination power among these patches given power constraints, and (3) robust performance for non-planar complex LOS geometries and improved NLOS localization and identification using our adaptive lighting algorithm. To validate these contributions, we demonstrate results using a physically-based renderer as well as a real working experimental setup in the lab. We publicly release our code, trained network models, and datasets at \url{https://github.com/sreenithy/AdaptiveLighting\_NLOS} for use by the research community. This paper combines the advantages of physics-based vision with deep learning, and shows how the choice of illumination can play a major role in the performance of NLOS imaging algorithms.

\section{Related Work}
\textbf{NLOS Imaging with Time-Resolved Detectors:} Kirmani et al. introduced ``looking around corners" using 5D time-resolved light transport~\cite{kirmani2009looking}. This was later realized using streak cameras and a femto-second laser to perform elliptic backpropagation~\cite{velten2012recovering, gupta2012reconstruction}. Researchers demonstrated NLOS reconstruction using continuous-wave time-of-flight sensors via optimization~\cite{heide2014diffuse} and NLOS tracking using array signal processing~\cite{kadambi2016occluded}. Single-photon avalanche diodes (SPADS)~\cite{Cova96} have been used for NLOS reconstruction~\cite{buttafava2015non,o2018confocal}. Advances in reconstruction algorithms for SPADs include accounting for partial occluders and surface normals~\cite{heide2017robust}, confocal NLOS reconstruction~\cite{o2018confocal}, geometric modeling~\cite{pediredla2017reconstructing}, backpropagation~\cite{arellano2017fast,la2018error}, space-carving~\cite{tsai2017geometry}, temporal focusing~\cite{SNLOS}, wave-based models in the frequency domain~\cite{waveNLOS}, and the identification of Fermat paths~\cite{xin2019theory}. Higher level applications such as detection and tracking have been shown~\cite{gariepy2016detection,chan2017non}. Recently, neural networks have been applied to time-resolved measurements for NLOS imaging~\cite{Caramazza,satat2017object}.

\textbf{NLOS Imaging without Time-Resolved Detectors:} Sen et al. transposed the light transport matrix to see an object in the camera's NLOS, although it is still in the projector's LOS~\cite{sen2005dual}. Accidental pinholes and pinspeck cameras can visualize the NLOS scene~\cite{torralba2014accidental}. Bouman et al. visualize 1D and 2D slices of the NLOS scene using a physical wall/corner~\cite{bouman2017turning}. Known occluders in the NLOS can help reconstruct the light field~\cite{baradad2018inferring} and hidden scene~\cite{thrampoulidis2018exploiting}. Smith et al. used laser speckle for tracking NLOS objects~\cite{smith2018tracking}. Similar to this paper, Klein et al. demonstrate a real-time tracking algorithm using radiosity~\cite{klein2016tracking}, Tancik et al. perform deep learning for conventional cameras for NLOS imaging~\cite{tancik2018data,tancik2018flash}, and Chen et al. recently show high quality reconstruction from RGB cameras using a convolutional neural network~\cite{steady-state}. However, our work differs from the previous works by introducing the idea of adaptive lighting to improve the performance of data-driven NLOS localization, and show this for non-planar LOS scenes.

\section{NLOS Localization and Identification with Deep Learning}

Our NLOS imaging scenario consists of a projector illuminating a spot on a LOS scene (not necessarily planar), and a camera captures an image of this spot. An illuminated spot on the wall undergoes at least three diffuse reflections or bounces as it travels from a point on the wall to hidden NLOS object and back to the wall, before being captured by the focused camera. This setup is present in the top left corner of Fig.~\ref{fig:problem}. The tasks we are interested in include localization, i.e. determining the $(x,y,z)$ location of an object's centroid, and identification of that object is present in the NLOS. In this paper, we assume a fixed surveillance scenario where information about the line-of-sight (LOS) is known, and images can be captured for training of neural networks. While these assumptions limit the practical application of this research (i.e. does not generalize to unseen locations or moving scenes), it allowed us tractability in which to investigate questions about localization, identification, and adaptive lighting.

\subsection{Dataset Rendering}
For deep learning, large datasets of images are typically needed to train the networks. However, collecting such large numbers of image data for a real NLOS scene would be prohibitive. Thus we utilize synthetic data to train our algorithms, and then fine-tune the networks on the real experimental dataset which is smaller in scale. We use dataset augmentation including flipping, cropping, translation, and rotation, to help the network be robust to changes in camera viewpoint and improve generalisation performance. For synthetic data generation, we use the physically-based renderer Mitsuba~\cite{Mitsuba}. We render scenes similar to those shown in Figure~\ref{fig:setup}(A), that include complex geometries, occlusions, colors and textures. In our experimental results, we show how our adaptive lighting method robustly improves performance on all these scenes.

We used four types of NLOS objects: a sphere (diameter 5cm), a human model (width 5.5 cm and height 17.5 cm), a cylinder (diameter 6 cm and height 8 cm), and the Stanford bunny (width 7.4 cm and height 5 cm). The illuminated spot was randomly positioned on the LOS wall. For the adaptive lightings described later in Section~\ref{sec:adaptivelighting}, we position the spot(s) to illuminate the patch(es) given by the optimization algorithm.

To quickly model image formation, we use the Instant Radiosity algorithm~\cite{keller1997instant} available in Mitsuba. This does not generate physically-realistic images as compared to using Monte Carlo integration for solving the rendering equation~\cite{kajiya1986rendering,veach1997robust}, but it enables extremely fast rendering speeds. Rendering one image of $64 \times 64$ resolution with 20,000 samples per pixel (spp) takes 18 seconds on a GeForce GTX1080 Ti GPU. This enabled dataset creation up to 100,000 images suitable for deep learning in a short amount of time, while rendering a full Monte Carlo path tracer at 20,000 spp takes 22 minutes per image. Previous work has shown that physically-realistic rendering using Monte Carlo compared to radiosity does lead to improved performance for computer vision tasks~\cite{zhang2016physically}. However, this paper localizes an object of width 5.5 cm with an MSE of 1.55 cm using radiosity images. We leave it to future work to show NLOS scenes where physically-realistic rendering is critical for learning performance. 

\subsection{Network Architectures}

\begin{figure}[t]
\setlength{\textfloatsep}{0.1\baselineskip plus 0.2\baselineskip minus 0.5\baselineskip}
\begin{center}
\includegraphics[width=1\columnwidth]{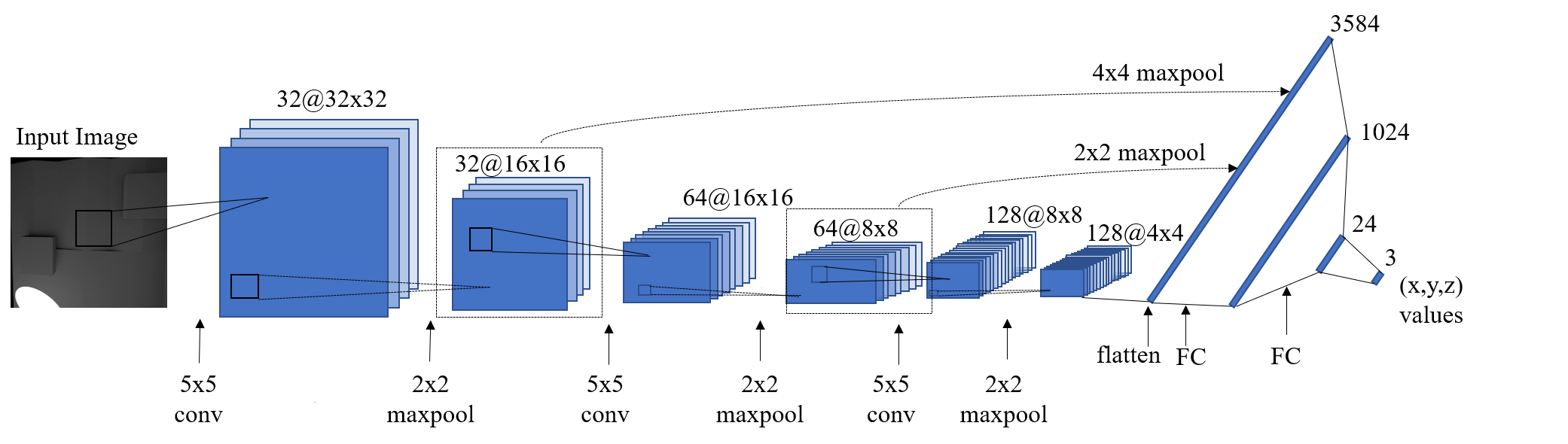}
\end{center}
  \caption{Proposed network architecture for NLOS localization which uses convolutional layers followed by fully connected layers to perform the end task. }
\label{fig:arch1}
\end{figure}

\begin{figure}[t]
\begin{center}
%\fbox{\rule{0pt}{2in} \rule{0.9\linewidth}{0pt}}
\includegraphics[width=1\columnwidth]{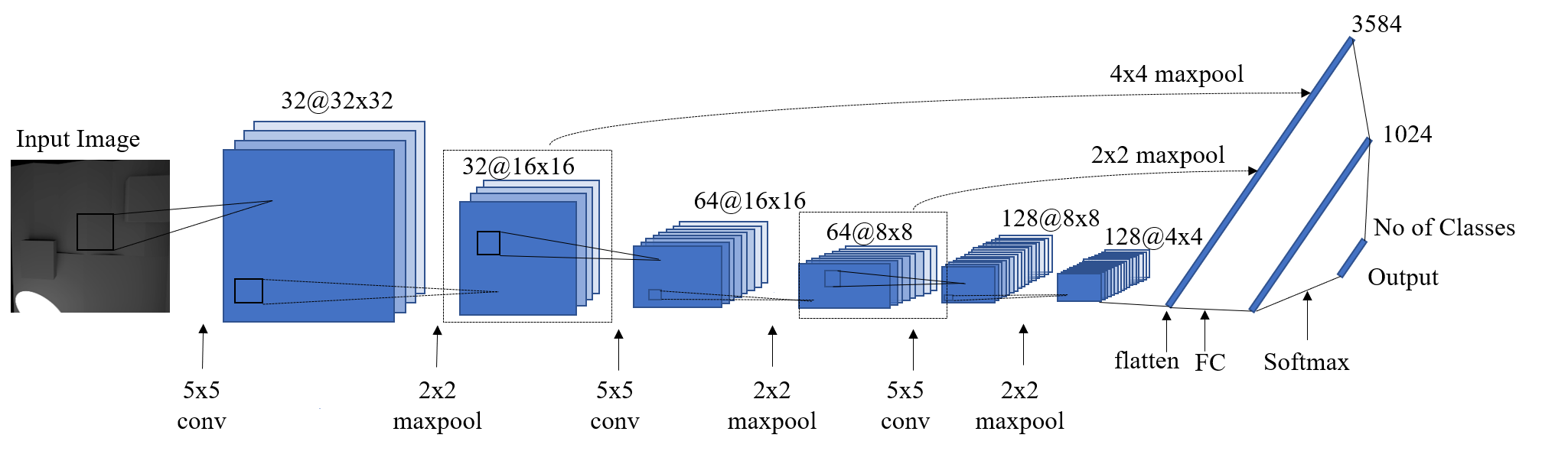}
\end{center}
   \caption{Proposed network architecture for NLOS Identification }
\label{fig:identify}
\end{figure}

Our network architectures are similar for localization and identification with a few changes to the last layers. The network architecture consists of five layers: three convolutional layers for feature extraction and either two fully connected layers for localization, or a fully connected layer and soft-max classifier for object identification. We use multi-scale features via skip connections similar to~\cite{sermanet2011traffic}. Convolution layers  are of size 5 $\times$ 5 with a stride of 1. Pooling layers  are of size 2 $\times$ 2. A ReLU is performed after each pooling layer and before each fully connected layer. The localization network has three outputs corresponding to the predicted $(x,y,z)$ centroid location. The identification network uses softmax for classification into object classes. Each network is trained with images with multiple NLOS objects but only one type of LOS scene/wall. The network architecture for localization is in Fig \ref{fig:arch1}, a similar network architecture for classification is shown in Fig \ref{fig:identify}. Note that we do not perform background subtraction or calibration with knowledge that there is no NLOS object in the scene for our images at testing/inference as compared to other works~\cite{klein2016tracking,tancik2018flash}. For implementation details including training times, epochs, learning rate, please see the supplemental material.

\section{Adaptive Lighting for Improved NLOS Localization and Identification}
\label{sec:adaptivelighting}
A central question we explore in this paper is what are the best lighting patterns in the LOS which maximize NLOS localization and identification? In other words, given a LOS scene with $N$ patches, what patches maximize photons that travel along NLOS light paths? To approach this problem, we use physics-based constraints on light transport given by radiosity~\cite{cohen2012radiosity,goral1984modeling,keller1997instant}. Radiosity describes the transfer of radiant energy between surfaces in the scene, and is calculated based on surface geometry and reflectances/emittances. We make the following modeling assumptions: (1) we have knowledge of the LOS geometry (i.e. via structured light or 3D scanning), (2) diffuse reflection and emittance only in our scene, and (3) only modeling up to third bounce light similar to assumptions made by Klein et al.~\cite{klein2016tracking}. We thus formulate an optimization to determine lighting that will maximize signal from the NLOS, and can handle complex LOS scenes with occlusions and non-planar geometries. This is a first step to having NLOS imaging work for complex LOS scenes in the real world.    

The total radiosity leaving a single patch will be its emission (i.e. if its a light source) along with the summation of all reflected radiosity from other patches~\cite{goral1984modeling}:
\begin{equation}\label{eq:1}
    B_{i}=E_{i}+\rho_{i} \sum_{j=1}^{N}F_{ij}B_{j}, \hspace{1cm}F_{ij}=\Big(\frac{cos \theta_i cos \theta_j}{\pi r^2} \Big)V(x_i,x_j) A_i\
\end{equation}
where $B_{i}$ is the outgoing radiosity of patch $i$, $E_{i}$ is the energy emitted out of the patch $i$, $\rho_{i}$ is the diffuse reflectance of the patch $i$, $F_{ij}$  is the form factor which determines how much light is transferred from one patch to another, and  $B_{j}$ is the outgoing radiosity of patch $j$. $F_{ij}$ depends on \(x_i \) as the central point on patch $i$, \(x_j \) as a central point on patch $j$, $r = ||x_i - x_j||$ as the distance between \(x_i\) and \(x_j\), \( \theta_i\) as the angle between normal \(N_i\) and vector $x_j - x_i$, \( \theta_j\) as the angle between normal \(N_j\) and vector $x_i - x_j$, and $V(x_i,x_j)$ as the visibility function of both patches. Note that this is an approximation of radiosity, and that radiosity form factors can be better calculated using hemicubes~\cite{cohen1985hemi}. However, this is a common approximation used in practice for NLOS imaging~\cite{heide2017robust,klein2016tracking}.

This visibility function accounts for occlusion, and is 1 if patch $i$ is visible from patch $j$, and 0 otherwise. We utilize basic raycasting to determine if two patches are visible to one another, which allows us to handle complex LOS scenes with occlusion. However, we are not robust to occlusion in the NLOS as this is difficult to determine from only LOS information. 

In the LOS scenes we consider, we include objects that have varying colors and material reflectance that include mixed diffuse ($k_d$) and specular ($k_s$) terms. However, for our algorithm, we only consider the $k_d$ term to formulate a radiosity optimization. Of course accounting for specularity may improve the performance of our adaptive lighting algorithm, but we found that we achieved significant improvement using diffuse modeling assumptions alone.

% Future work can extend our framework to specular or even more complex BRDFs such as the technique demonstrated in~\cite{steady-state}. 

\subsection{Maximizing NLOS Radiosity}
We now present our adaptive lighting optimization algorithm based on radiosity. Please refer the supplemental material for a more complete derivation for the light paths that are chosen for optimizing over. Using the above radiosity equation, we can calculate the radiosity for LOS light paths and NLOS light paths for three bounce light. If we seek to illuminate $m$ patches out of $N$ total LOS patches, we wish to maximize the following:
\begin{equation}\label{eq:multipatch}
    \argmax_{[1, \dots, m] \subseteq [1,\dots,N]} \sum_{i=1}^{m} B_i^{NLOS}
\end{equation} where $B_i^{NLOS}$ is the NLOS radiosity from patch $i$. We can determine $B_i^{NLOS}$ by summing all three-bounce light paths that hit both the patch $i$ and the NLOS voxel (see supplemental material for full derivation and algorithm). We return the top $1,\dots,m$ patches which maximize the NLOS radiosity\footnote{Here, the indices of these patches are relabeled to be $1,\dots,m$ for notation. The optimization returns $m$ indices corresponding to the $m$ patches that is a random subset of the $N$ total patches.}. In the supplemental material, we discuss why adaptive lighting focuses on patches as opposed to spatially-varying lighting, which we show is not optimal because it spreads out power over a large region rather than concentrating photons on a few key patches. 

\textbf{Distributed Light Algorithm:} The real advantage of adaptive lighting is when the imaging system has tight power constraints for its illumination. This is common for embedded and mobile computer vision platforms such as robotics and unmanned aerial vehicles. New energy-efficient active illumination also exploits illuminating and capturing only salient light paths to extract signals of interest~\cite{o2015homogeneous,achar2017epipolar,kubo2018acquiring, ueda2019slope,wang2018programmable}. In this vein, we develop a \textit{distributive} algorithm which tells us not only what are the optimal patches to shine light, but what fraction of power from our power budget should we shine on each patch. 

We formulate a joint optimization over the $m$ optimal patches and the illumination radiant exitances $I_i, i=1,\dots,M$ for each patch. This is given by the following equation:
\begin{equation}\label{eq:redistributed}
\argmax_{\substack{[1,\dots,m]\subseteq [1,\dots,N] \\ I_{1},\dots,I_{m}}} \sum_{i=1}^{m} B_{i}^{NLOS}(I_i) \text{ such that } \sum_{i=1}^{m} I_i = T, \forall I_i \leq I_o.
\end{equation}
Here, $B_{i}^{NLOS}(I_i)$ is the radiosity for patch $i$ given illumination radiant exitance $I_i$, $T$ is the total illumination power constraint, and each radiant exitance cannot exceed a maximum $I_o$. 

We use the Sequential Least Squares Programming algorithm available in SciPy library~\cite{scipy} to solve our optimization problems. The optimization treats the indices as continuous variables to utilize conventional solvers (as opposed to combinatorial optimization), and we find the solutions converged to valid integer values nevertheless. The full optimization for adaptive lighting takes between 1-2 minutes to run on a conventional CPU. 

\textbf{Training and Inference using Adaptive Lighting:}
We describe the training and inference procedure for our adaptive lighting algorithm. This is summarized in Figure~\ref{fig:problem}. To determine adaptive lighting patterns for the NLOS, we divide the NLOS region into voxels in order to calculate the optimization that maximizes radiosity back from this voxel region. Then for our dataset, we train the network with an object contained in NLOS voxel $i$ with the associated adaptive lighting given by our algorithm for the same voxel. We are thus training the network to recognize these adaptive lightings per voxel class to maximally extract NLOS information. At inference, we present $N$ adaptive lightings, each corresponding to an associated possible $N$ NLOS voxels where the object could be located. We modify the final layer of the network to have a softmax output to give probabilities per voxel. We run all these $N$ images through forward propagation of the network and save this softmax output of each image. We then take the maximum confident probability across all $N$ images as the location of the NLOS object. 
\section{Adaptive Lighting for Improved NLOS Localization and Identification}
\label{sec:adaptivelighting}
A central question we explore in this paper is what are the best lighting patterns in the LOS which maximize NLOS localization and identification? In other words, given a LOS scene with $N$ patches, what patches maximize photons that travel along NLOS light paths? To approach this problem, we use physics-based constraints on light transport given by radiosity~\cite{cohen2012radiosity,goral1984modeling,keller1997instant}. Radiosity describes the transfer of radiant energy between surfaces in the scene, and is calculated based on surface geometry and reflectances/emittances. We make the following modeling assumptions: (1) we have knowledge of the LOS geometry (i.e. via structured light or 3D scanning), (2) diffuse reflection and emittance only in our scene, and (3) only modeling up to third bounce light similar to assumptions made by Klein et al.~\cite{klein2016tracking}. We thus formulate an optimization to determine lighting that will maximize signal from the NLOS, and can handle complex LOS scenes with occlusions and non-planar geometries. This is a first step to having NLOS imaging work for complex LOS scenes in the real world.    

The total radiosity leaving a single patch will be its emission (i.e. if its a light source) along with the summation of all reflected radiosity from other patches~\cite{goral1984modeling}:
\begin{equation}\label{eq:1}
    B_{i}=E_{i}+\rho_{i} \sum_{j=1}^{N}F_{ij}B_{j}, \hspace{1cm}F_{ij}=\Big(\frac{cos \theta_i cos \theta_j}{\pi r^2} \Big)V(x_i,x_j) A_i\
\end{equation}
where $B_{i}$ is the outgoing radiosity of patch $i$, $E_{i}$ is the energy emitted out of the patch $i$, $\rho_{i}$ is the diffuse reflectance of the patch $i$, $F_{ij}$  is the form factor which determines how much light is transferred from one patch to another, and  $B_{j}$ is the outgoing radiosity of patch $j$. $F_{ij}$ depends on \(x_i \) as the central point on patch $i$, \(x_j \) as a central point on patch $j$, $r = ||x_i - x_j||$ as the distance between \(x_i\) and \(x_j\), \( \theta_i\) as the angle between normal \(N_i\) and vector $x_j - x_i$, \( \theta_j\) as the angle between normal \(N_j\) and vector $x_i - x_j$, and $V(x_i,x_j)$ as the visibility function of both patches. Note that this is an approximation of radiosity, and that radiosity form factors can be better calculated using hemicubes~\cite{cohen1985hemi}. However, this is a common approximation used in practice for NLOS imaging~\cite{heide2017robust,klein2016tracking}.

This visibility function accounts for occlusion, and is 1 if patch $i$ is visible from patch $j$, and 0 otherwise. We utilize basic raycasting to determine if two patches are visible to one another, which allows us to handle complex LOS scenes with occlusion. However, we are not robust to occlusion in the NLOS as this is difficult to determine from only LOS information. 

In the LOS scenes we consider, we include objects that have varying colors and material reflectance that include mixed diffuse ($k_d$) and specular ($k_s$) terms. However, for our algorithm, we only consider the $k_d$ term to formulate a radiosity optimization. Of course accounting for specularity may improve the performance of our adaptive lighting algorithm, but we found that we achieved significant improvement using diffuse modeling assumptions alone.

% Future work can extend our framework to specular or even more complex BRDFs such as the technique demonstrated in~\cite{steady-state}. 

\subsection{Maximizing NLOS Radiosity}
We now present our adaptive lighting optimization algorithm based on radiosity. Please refer the supplemental material for a more complete derivation for the light paths that are chosen for optimizing over. Using the above radiosity equation, we can calculate the radiosity for LOS light paths and NLOS light paths for three bounce light. If we seek to illuminate $m$ patches out of $N$ total LOS patches, we wish to maximize the following:
\begin{equation}\label{eq:multipatch}
    \argmax_{[1, \dots, m] \subseteq [1,\dots,N]} \sum_{i=1}^{m} B_i^{NLOS}
\end{equation} where $B_i^{NLOS}$ is the NLOS radiosity from patch $i$. We can determine $B_i^{NLOS}$ by summing all three-bounce light paths that hit both the patch $i$ and the NLOS voxel (see supplemental material for full derivation and algorithm). We return the top $1,\dots,m$ patches which maximize the NLOS radiosity\footnote{Here, the indices of these patches are relabeled to be $1,\dots,m$ for notation. The optimization returns $m$ indices corresponding to the $m$ patches that is a random subset of the $N$ total patches.}. In the supplemental material, we discuss why adaptive lighting focuses on patches as opposed to spatially-varying lighting, which we show is not optimal because it spreads out power over a large region rather than concentrating photons on a few key patches. 

\textbf{Distributed Light Algorithm:} The real advantage of adaptive lighting is when the imaging system has tight power constraints for its illumination. This is common for embedded and mobile computer vision platforms such as robotics and unmanned aerial vehicles. New energy-efficient active illumination also exploits illuminating and capturing only salient light paths to extract signals of interest~\cite{o2015homogeneous,achar2017epipolar,kubo2018acquiring, ueda2019slope,wang2018programmable}. In this vein, we develop a \textit{distributive} algorithm which tells us not only what are the optimal patches to shine light, but what fraction of power from our power budget should we shine on each patch. 

We formulate a joint optimization over the $m$ optimal patches and the illumination radiant exitances $I_i, i=1,\dots,M$ for each patch. This is given by the following equation:
\begin{equation}\label{eq:redistributed}
\argmax_{\substack{[1,\dots,m]\subseteq [1,\dots,N] \\ I_{1},\dots,I_{m}}} \sum_{i=1}^{m} B_{i}^{NLOS}(I_i) \text{ such that } \sum_{i=1}^{m} I_i = T, \forall I_i \leq I_o.
\end{equation}
Here, $B_{i}^{NLOS}(I_i)$ is the radiosity for patch $i$ given illumination radiant exitance $I_i$, $T$ is the total illumination power constraint, and each radiant exitance cannot exceed a maximum $I_o$. 

We use the Sequential Least Squares Programming algorithm available in SciPy library~\cite{scipy} to solve our optimization problems. The optimization treats the indices as continuous variables to utilize conventional solvers (as opposed to combinatorial optimization), and we find the solutions converged to valid integer values nevertheless. The full optimization for adaptive lighting takes between 1-2 minutes to run on a conventional CPU. 

\textbf{Training and Inference using Adaptive Lighting:}
We describe the training and inference procedure for our adaptive lighting algorithm. This is summarized in Figure~\ref{fig:problem}. To determine adaptive lighting patterns for the NLOS, we divide the NLOS region into voxels in order to calculate the optimization that maximizes radiosity back from this voxel region. Then for our dataset, we train the network with an object contained in NLOS voxel $i$ with the associated adaptive lighting given by our algorithm for the same voxel. We are thus training the network to recognize these adaptive lightings per voxel class to maximally extract NLOS information. At inference, we present $N$ adaptive lightings, each corresponding to an associated possible $N$ NLOS voxels where the object could be located. We modify the final layer of the network to have a softmax output to give probabilities per voxel. We run all these $N$ images through forward propagation of the network and save this softmax output of each image. We then take the maximum confident probability across all $N$ images as the location of the NLOS object. 

\section{Experimental Results}
\label{sec:simulatedresults}

\begin{figure}
\begin{center}
\includegraphics[width=1\columnwidth]{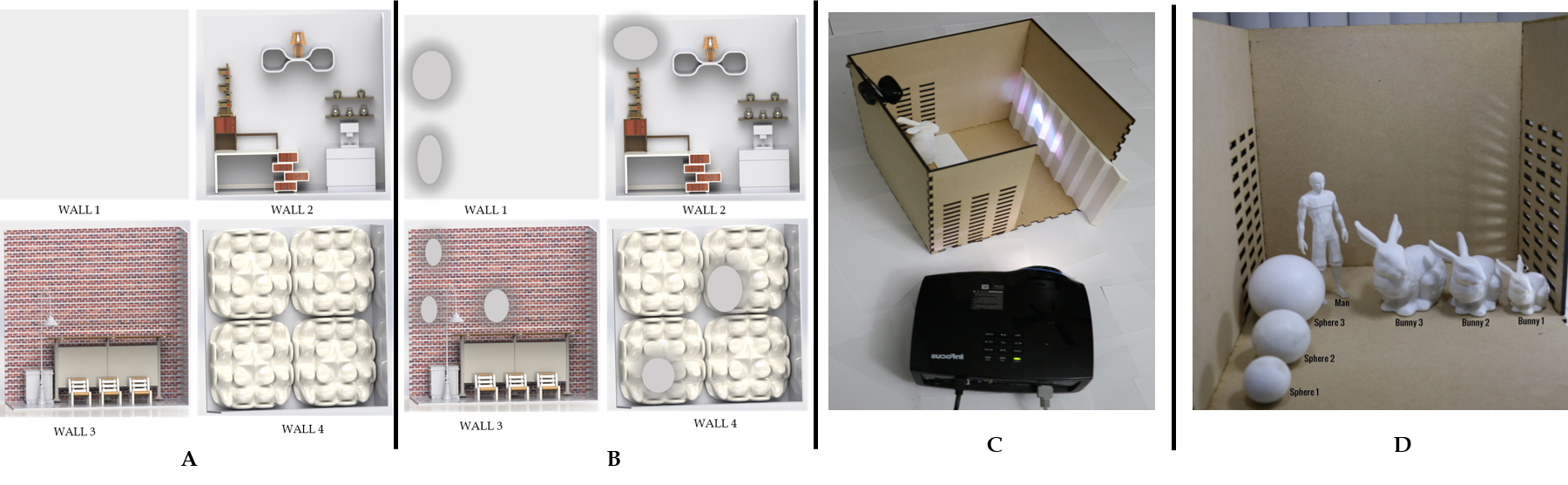}
\end{center}
  \caption{A) The various LOS scenes used for experiments; (B) The LOS scenes with optimal adaptive lighting patterns visualized; (C) Real experimental prototype in the lab; (D) The 3D printed objects used in real experiments. }
\label{fig:setup}
\end{figure}

\textbf{Simulated Results:}
In Table~\ref{tab:table1}, we display the results for simulated data for four types of LOS walls for the localization and identification of the 4 objects. For each task, we perform adaptive lighting to select the optimal patch ($m=1$) to illuminate, and the 2nd and 3rd choice correspond to the subsequent patches returned by the adaptive lighting optimization. As can be determined from the table, the adaptive optimal patch had an average mean-squared error (MSE) of 2.09 cm, while the second and third chosen patches had an average MSE of 3.60 cm and 5.08 cm respectively. Thus adaptive lighting improved localization on average by 1.51 cm across all experiments. In addition, we tested object identification in the NLOS. This was a four class object identification with an additional fifth class of ``No object" if the softmax score of the classifier was below $50\%$. Average identification of the NLOS object calculated across the four walls is $91.02\%$ for the adaptive lighting, while the next best illuminated patch averaged $82.28\%$. 

\textit{Network Generalisation:} Our network generalizes across changes in camera viewpoint at testing, since we performed data augmentation during training of this nature. We tested rotating a camera a full 360 degrees in increments of 45 degrees at test time (for a NLOS sphere using LOS Wall 2). This incurred localization errors that varied up to 0.75 cm. Similarly, for translation of the camera position along the x,y,z axis, we only saw variation up to 0.4 cm. Thus our dataset augmentation yielded generalisation to camera motion at test time. 

However, our network does not generalize as well to NLOS objects its never seen before. Using new NLOS object of the Stanford dragon (length of 6 cm, width of 3.5 cm and height of 5 cm) resulted in 8.02 cm localization error, and the Happy Buddha status (width of 5 cm and height of 15 cm) resulted in 7.18 cm localization error. It remains future research to make the networks more robust to unknown objects or simply to add enough data for the network to generalize.

\textit{Distributed Lighting:} Also in Table~\ref{tab:table1}, we present the distributed lighting algorithm results that jointly optimizes over patch selection and illumination power(Equation~\ref{eq:redistributed}). To compare, we present two baseline methods: Method 1 only optimizes patch selection using the adaptive lighting algorithm and distributes power evenly among these patches, while Method 2 neither optimizes the distribution of power nor the selection of patches. Our simulated experiment is for two objects (Man and Bunny) with LOS Wall 3, for two $(m=2)$ illumination patches. Note how we get improvements of $40\%(3.25\text{ cm} \to 2.17\text{ cm} \text{ average})$ improvement of jointly optimizing over Method 1, and as high as $ 65\% (5.22\text{ cm} \to 2.17\text{ cm} \text{ average})$ improvement over Method 2.

\begin{figure}
\begin{center}
\includegraphics[width=12.5cm,height=5cm]{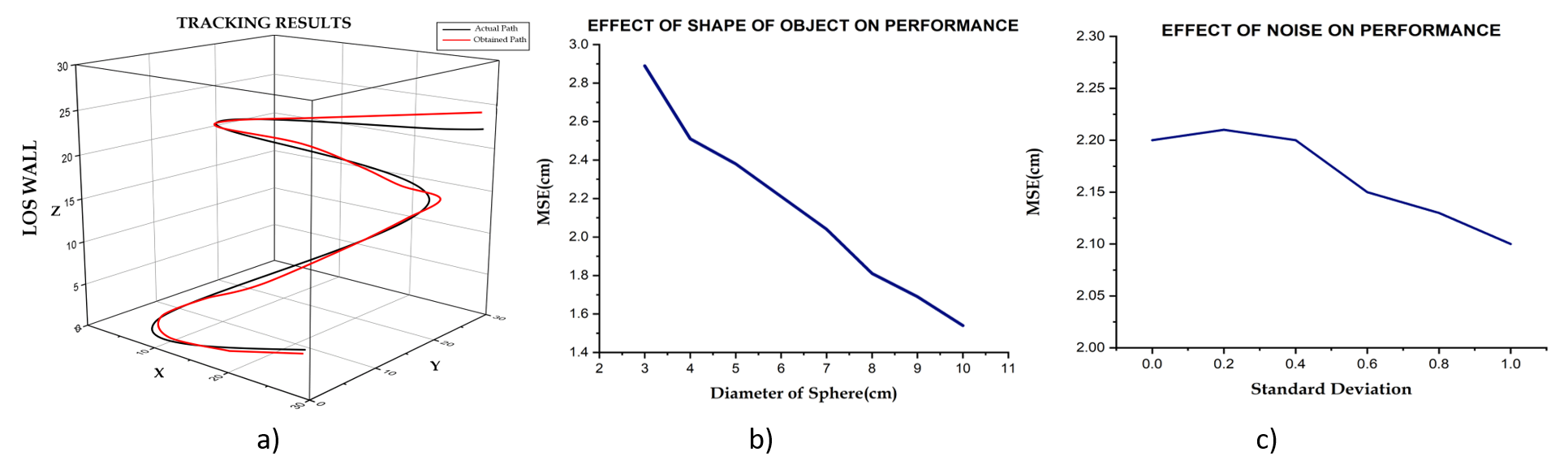}
\end{center}
  \caption{a) Tracking results for a sphere of diameter 5 cm for simulated data. The black trajectory denotes the actual path taken by object and red denotes the predicted position of the object by using the network proposed in Fig~\ref{fig:arch1}; b)Effect of shape of object on the MSE. As the diameter of the sphere increases, the MSE reduces, giving better accuracy in localization; c) Effect of adding noise to the dataset while training.}
\label{fig:ablative}
\end{figure}

\textit{Tracking and Ablation Studies:} In Figure~\ref{fig:ablative}, we show tracking results for our NLOS localization, as well as ablation studies for the effects of object size and noise. Figure~\ref{fig:ablative}(a) depicts the tracking results of a sphere, with the black trajectory denoting the ground truth and red denoting the network localization. When the sphere is within approximately 10 cm from the LOS wall(one-third of the LOS wall length), the localization error is smaller and it tends to increase progressively as the object moves farther away from the LOS wall. This is due to the loss of light due to the radiometric falloff as the NLOS object moves farther away, conforming to our intuitions. 

For object size, we varied the diameter of a simulated sphere in the NLOS, and noticed how localization performance improves as the object gets bigger. This is intuitive since a larger object yields more reflected light back to the LOS which can be extracted by the network. Noise is usually added to make the model more robust to changes~\cite{greff2017lstm}. We additionally study the effect of adding noise to the training examples. In Figure~\ref{fig:ablative}(c), we add Gaussian noise to all training examples, and report the MSE performance of the resulting network. Note that the performance stays relatively flat, showing that the model is relatively robust to noise (this may be because the renderer itself has noise that helps regularize the network). In the supplemental material, we also perform saliency analysis to determine what parts of the LOS scene is salient for the NLOS localization task.

\begin{table} \footnotesize
        \begin{center}
        \begin{tabular}{|c|c|c|c|c|c|c|}%{ |M{20mm}|M{12mm}|M{9mm}|M{8mm}|M{8mm}| }
       
         \hline
         Wall &Object & \multicolumn{3}{c|}{Localization with Adaptive Lighting}  & \multicolumn{2}{c|}{Identification with Adaptive Lighting}\\
      \cline{3-7}
    &&Optimal Patch& 2nd Patch& 3rd Patch&Optimal Patch&2nd Patch\\  
\cline{1-7}
     1 &SPHERE &2.19 cm&3.26 cm& 4.89 cm  &92.6\%&81.7\%\\
     1 &BUNNY & 1.89 cm& 2.54 cm& 5.93 cm &92.8\%&82.3\%\\
     1 &MAN &1.50 cm& 2.30 cm& 3.91 cm &93.3\%&82.1\% \\
     1 &CYLINDER &1.52 cm& 2.94 cm& 3.75 cm & 93.6\%&84.2\%\\
     1&NO OBJECT&-&-&-&98.5\%&96.3\%\\
     2 &SPHERE &2.34 cm& 3.81 cm& 5.23 cm &89.9\%&79.1\%\\
     2 &BUNNY &2.31 cm &3.37 cm& 4.91 cm &90.3\%&84.9\% \\
     2 &MAN &1.56 cm& 3.09 cm& 4.87 cm &91.4\%&81.3\% \\
     2 &CYLINDER &1.98 cm&3.92 cm& 4.11 cm& 91.0\%&83.5\%\\
     2&NO OBJECT&-&-&-&98.1\%&94.6\%\\
     3 &SPHERE &2.90 cm&4.9 cm& 6.15 cm&86.4\%&76.1\%\\
     3 &BUNNY &2.81 cm&4.17 cm& 5.85 cm&87.8\%&78.3\%\\
     3 &MAN &1.92 cm&3.98 cm& 5.02 cm &88.1\%&79.2\%\\
     3 &CYLINDER &2.30 cm&3.26 cm& 5.68 cm&89.3\%&80.6\%\\
     3&NO OBJECT&-&-&-&97.6\%&93.0\%\\
     4 &SPHERE &2.98 cm&4.51 cm& 5.81 cm&84.6\%&71.8\% \\
     4& BUNNY& 1.87 cm&4.20 cm& 5.96 cm &85.9\%&73.4\%\\
     4 &MAN &1.63 cm&3.43 cm& 4.56 cm  &86.3\%&75.6\% \\ 
     4& CYLINDER &1.71 cm&3.87 cm& 4.69 cm &87.1\%&78.3\%\\
     4&NO OBJECT&-&-&-&95.8\%&89.3\%\\
    \hline
    \multicolumn{7}{|c|}{Distributed Lighting}  \\ \hline \hline
   \multicolumn{2}{|c|}{Object}& \multicolumn{3}{c|}{Optimal lighting using distributed lighting}& Method 1& Method 2\\
    \hline
     \multicolumn{2}{|c|}{MAN}& \multicolumn{3}{c|}{1.56 cm} &2.45 cm& 4.32 cm\\
     \multicolumn{2}{|c|}{BUNNY} & \multicolumn{3}{c|}{2.36 cm} &3.19 cm& 5.78 cm\\
     \hline
        \end{tabular}
           \end{center}
        \caption{Simulated Results for 4 different walls and 4 objects, including the performance improvement due to adaptive lighting for the optimal patch versus 2nd and 3rd patches returned by the optimization. Distributed lighting optimizes both patch selection and power distribution for $m=2$ patches, while Method 1 only performs adaptive patch selection with equal distribution of power and Method 2 does not optimize either the illumination or patches.  
}
        \label{tab:table1}
    \end{table}

    % Don't include this in the table caption
            % The average MSE across all the four objects for adaptive localization method is 2.09cm. And using the second best optimal patch illumination is 3.60, third-best illumination condition is 5.08cm. The adaptive identification averages at  91.02\%, non-adaptive identification at 82.28\%. While using the distributed method for illuminating two different patches under a power constraint localized objects with an  average MSE of 2.17cm against 3.25cm when patches are optimally chosen and illumination power is not optimized and 5.22cm when neither illumination nor patch selection is optimized
    
% \begin{figure}

% \begin{center}
% \includegraphics[width=1\columnwidth]{IMAGES/all.png}
% \end{center}
%   \caption{a) Tracking Results for a sphere of diameter 5cm for simulated data. The black trajectory denotes the actual path taken by object and red denotes the predicted position of the object by using the network proposed in Fig~\ref{fig:arch1}; b)Effect of shape of object on the MSE. As the diameter of the sphere increases, the MSE reduces, giving better accuracy in localization; c) Effect of adding noise to the dataset while training.}
% \label{fig:ablative}
% \end{figure}    

%     \begin{figure}
% % \begin{center}
% \includegraphics[width=0.32\columnwidth]{IMAGES/TRACKNEW1.png}
% \includegraphics[width=0.32\columnwidth]{graph4new.jpg}
% \includegraphics[width=0.32\columnwidth]{graph10new.jpg}
% % \end{center}
% \vspace{2mm}
%   \caption{\sj{Write a caption for this}Object Tracking, size of object, noise}
% \label{fig:ablative}
% \end{figure}

\textbf{Real Experimental Results:}
For our hardware setup, we constructed a physical NLOS imaging prototype (35.6 cm $\times$ 35.6 cm $\times$ 35.6 cm dimensions) to validate our proposed method in the real world, as shown in Figure~\ref{fig:setup}(C). The LOS scene used is a variation of LOS Wall 2 in that figure, and consisted of black, white, and grey colors and coarse texture given by the 3D printer. Objects were also 3D printed. Further implementation details are provided in the supplemental material.

In Table~\ref{tab:table2}, we present the results from our real hardware prototype. The adaptive lighting for a single patch resulted in an average MSE of 1.97 cm in localization, while the next best-patch had average error of 2.95 cm. The values reported in Table~\ref{tab:table2} is for a sphere (diameter of 5 cm), cylinder(radius of 3 cm and height of 8 cm), a bunny (width of 7.4 cm and height of 5 cm), and a man silhouette (width of 5.5cm and height of 17.5cm). For identification, we have similar improvements in accuracy due to optimal choice of lighting (87.1\%) as opposed to the next-best choice (78.3\%). 

For the distributed method, we run our algorithm to optimize for two patches simultaneously in terms of location and illumination power. The adaptive algorithm achieves a MSE of 1.78 cm for the man and 2.55 cm for the bunny, 2.63 cm for the ball, and 2.41 cm for the cylinder; all of which outperform both Method 1 and  Method 2. Our real results help validate the idea that adaptive lighting can help improve NLOS localization and identification in a real world scene.

 \begin{table}\footnotesize
        \begin{center}
        \begin{tabular}{|c|c|c|c|c|}%{ |M{20mm}|M{12mm}|M{9mm}|M{8mm}|M{8mm}| }
       
         \hline
         Object & \multicolumn{2}{c|}{Localization with Adaptive Lighting}  & \multicolumn{2}{c|}{Identification with Adaptive Lighting}\\
      \cline{2-5}
    &Optimal Patch&2nd Patch&Optimal Patch&2nd Patch\\  
\cline{2-5}
      MAN &1.55 cm &2.41 cm&90.1\%&81.9\%\\
      CYLINDER&2.07 cm&3.12 cm&85.0\%&75.8\%\\

BUNNY &1.89 cm &2.82 cm&87.5\%&78.6\%\\

BALL  &2.38 cm &3.45 cm&85.8\%&77.2\%\\
    \hline
    \multicolumn{5}{|c|}{Distributed Lighting}  \\ \hline \hline
   \multicolumn{1}{|c|}{Object}& \multicolumn{2}{c|}{Optimal Lighting}& Method 1& Method 2\\
    \hline
     \multicolumn{1}{|c|}{MAN}& \multicolumn{2}{c|}{1.78 cm} &2.97 cm& 5.01 cm\\
      \multicolumn{1}{|c|}{CYLINDER}& \multicolumn{2}{c|}{2.41 cm} &3.28 cm& 6.16 cm\\
     \multicolumn{1}{|c|}{BUNNY} & \multicolumn{2}{c|}{2.55 cm} &3.53 cm& 5.43 cm\\
     \multicolumn{1}{|c|}{BALL}& \multicolumn{2}{c|}{2.63 cm} &4.10 cm& 5.69 cm\\
     \hline
        \end{tabular}
           \end{center}
        \caption{ Real data results using our hardware prototype for four objects trained on the complex LOS wall 2. Note that our optimization algorithms consistently give better localization and identification performance as compared to the next-best patch illumination.}
        \label{tab:table2}
    \end{table}

% !TEX root = main.tex
\section{Discussion}

Our results provide several points of discussion, as well as further implications for adaptive lighting methods. We note the effectiveness of deep learning methods, even with fairly conventional CNN architectures, at being able to extract information from the NLOS light paths and perform localization. In our best real data result from our hardware prototype, we localize the centroid of an object of width 5.5 cm and height 17.5 cm to an MSE of 1.55 cm using our adaptive lighting method. This outperforms state-of-the-art results reported in the literature including 1.7 cm localization error for a 12 cm wide object (in 2D) from~\cite{tancik2018flash} and 6.1 cm localization error for a 10 cm wide object(in 3D) from~\cite{klein2016tracking}. Further, to the best of our knowledge, our paper is the first method to handle non-planar LOS scenes for NLOS imaging. Yet we caution the reader that it is difficult to make a fair comparison between different algorithms/methods as the objects, simulators, and training data used are not standardized. Recently, a quantitative benchmark for NLOS imaging was introduced that could potentially help in this regard~\cite{klein2018quantitative}. To further these efforts, we publicly release our code and dataset at this Github link: \url{https://github.com/sreenithy/AdaptiveLighting\_NLOS} for the research community.

There are several limitations for our research approach. It is unclear whether our method can perform full 3D NLOS object reconstruction with a conventional camera and projector. We showed that while our networks generalize with respect to camera motion, they are not tolerant to unknown NLOS objects or if the LOS scene changes. Finally, we have not considered occlusions in the NLOS that limits our modeling techniques for adaptive lighting. 

There are several avenues for future work. More physically-realistic light transport modeling, including specular light paths, would enable better adaptive lighting and better data sets for training. We chose to optimize radiosity to boost the number of photons returning from the NLOS as a first step towards adaptive lighting. However, a full end-to-end network which jointly optimizes the lighting patterns with the inverse NLOS task can probably yield superior results. New architecture designs for our CNNs could potentially improve their performance and generalizability, including incorporating generative adversarial networks or performing joint localization and identification. Finally, the method could be engineered for real-time localization/tracking at inference by accelerating the radiosity optimization with techniques such as cached shadow maps implemented on a GPU. However, we hope that this paper is the first step towards intelligent lighting for NLOS imaging in the future.

\vspace{0.2cm}
\textbf{Acknowledgments:} We would like to thank Srinivasa Narasimhan, Kyros Kutulakos, and Ioannis Gkioulekas for initial discussions. Shenbagaraj Kannapiran from ASU helped render illustrations used in this paper. The AME Fabrication lab also assisted with 3D printing, scene setup, and lending the projector for our hardware prototype. This work was partially supported by the Defense Advanced Research Projects Agency (REVEAL Grant HR00111620021) to S. Jayasuriya, as well as joint support from both the Herberger Research Initiative in the Herberger Institute for Design and the Arts (HIDA) and the Fulton Schools of Engineering (FSE) at ASU to S. Chandran and S. Jayasuriya. This work was also supported by a hardware GPU donation by NVIDIA.

%\bibliography{egbib}
%\bibliography{egbib}

\newpage

\section{Supplemental Material}
In this supplemental material, we present additional material concerning the derivation of our adaptive lighting algorithm based on radiosity, implementation details for our experiments, and a small analysis of visual saliency in our network performance. 

\subsection{Adaptive Lighting}

In the main paper, we presented our adaptive lighting algorithm that optimized over $B_{NLOS}$. In this section, we present the complete analytic derivations for radiosity for three bounce light used in that algorithm. We mainly follow the approach of Klein et al.~\cite{klein2016tracking} in calculating our radiosities.

Let \(S_1\),\(S_2\)....\(S_N\) be the $N$ patches on the reflective LOS wall, light source denoted as \(p\), camera denoted as \(C\) and the  NLOS patch denoted as \(NLOS\). To calculate the radiosity along a ray for three bounce light, we must first calculate its first and second bounces.

\paragraph{First Bounce (LOS): }\begin{equation*} p \Rightarrow S_i \Rightarrow C 
\quad \forall i \in \{1,N\}.\end{equation*}

When light travels from the source to a diffuse wall and bounces back to the camera, the associated radiosity is given as the product of the reflectance of the surface $\rho_i$, the radiosity of the incident light $B_p$, and the form factor $F_{ip}$  between the $p$ and the $i$th patch, and the visibility term $V_i$~\cite{cohen2012radiosity,keller1997instant}:
\begin{equation} \label{eq:1}
B_{first}=B_i=\rho_iB_pF_{ip}V_i,
\quad \forall i \in \{1,N\}.\end{equation}
The form factor calculates how much light is transferred from one patch to another. Since the wall is divided into $N$ patches, the first bounce radiosity associated with all the $N$ patches is calculated. It takes into account the distance between the surfaces, computed as the distance between the center of each of the surfaces, and their orientation in space relative to each other, computed as the angle between each surface's normal vector and a vector drawn from the center of one surface to the center of the other surface: 
\begin{equation} \label{eq:2}
F_{ij}=\frac{\cos\theta_j \cos\theta_i}{\pi(x_j-x_i)^2}\cdot A_i
\end{equation} 

A visual depiction is shown in Figure 1. However the above equation does not account for  occlusion between the two patches. This is accounted by the visibilty term $V(i, j)$:
$$
   V_i(\vec{x_a},\vec{x_b},\vec{N_a})= 
\begin{cases}0,& \text{if } k>=\frac{\pi}{2} \: and \: k<=\frac{3\pi}{2} \\
    1,              & \text{otherwise}
\end{cases}
$$
where $k=(\vec{x_a}-\vec{x_b})\cdot \vec{N_a}.$

 Substituting form factor and visibility terms into \eqref{eq:1}, we get the following expression for first bounce radiosity:

\begin{equation} \label{eq:3}
B_{first}=B_i=\rho_i\cdot B_p \cdot \Big(  \frac{\cos\theta_p \cos\theta_i}{\pi(x_p-x_i)^2}\Big )\cdot A_i \cdot V_i, \quad\forall i \in \{1,N\}.
\end{equation}

\paragraph{ Second Bounce (LOS): }\begin{equation*}
p \Rightarrow S_i \Rightarrow S_j  \Rightarrow C   \end{equation*} for $\forall i \in \{1,N\}, \forall j \in \{1,N\}, j\neq i.$ In this case, light from the illuminating source hits the diffuse wall and the gets reflected to another patch on the diffuse wall. This can be viewed as the light taking two bounces and containing only LOS scene information when it reaches the camera. Using the radiosity calculated from Equation(~\ref{eq:3}) as the radiosity illuminating a second bounce patch, we get the expression for \(B_{i}\): 
\begin{equation} \label{eq:4}
B_j=\rho_jB_iF_{ji}V_j
\end{equation}
$$\forall i \in  \{1,N\}, \forall j \in  \{1,N\},  j \neq i.$$

%\[\: \: \:\forall \:j\: \epsilon \: \{1,N\} but \neq i\]

\paragraph{Third Bounce:}
For third bounces, we now have two subcases: when the ray only interacts with LOS patches, and when the ray interacts with the NLOS patch. We treat each case separately in our derivations.
\newline
\textbf{LOS Condition}
\[ p \Rightarrow S_i \Rightarrow S_j \Rightarrow S_k \Rightarrow C, \]
\[\forall i \in \{1,N\}, 
\forall j \in  \{1,N\}, j\neq i, \forall k \in  \{1,N\}, k \neq j.\]

Consider the scenario where the light after bouncing off two diffuse wall patches strikes another diffuse wall patch. This can be viewed as the light taking three bounces and containing only LOS scene information when it reaches the camera. 
Using the radiosity calculated from Equation~\eqref{eq:4} as the radiosity illuminating a third bounce patch,  we get the expression for \(B_{k}\): 

\begin{equation} \label{eq:5}
B_k=\rho_kB_jF_{kj}V_k
\end{equation}

\textbf{NLOS Condition}
\[ p \Rightarrow S_i \Rightarrow NLOS \Rightarrow S_k \Rightarrow C \]

In the scenario where, after light after undergoing first bounce LOS reaches the NLOS scene and then bounces to a diffuse wall patch beore reaching the camera, we can view that as three bounce NLOS light. After the first bounce, the incident radiosity is given by the condition Equation~\eqref{eq:3}. 
\begin{equation} \label{eq:6}
B_{n}= \rho_{n}B_iF_{ni}V_{n}  
\end{equation}
\[ \forall \:i\: \in \: \{1,N\}, \forall \:n\: \in \: \{1,N\} \]
Using Equation~\eqref{eq:6} as the radiosity of the light reaching the diffuse wall patch, we obtain the third bounce as below:
\begin{equation} \label{eq:7}
B_{k}= \rho_{k}B_{n}F_{kn}V_{k},     
\end{equation}
\[  \forall \:k\:\in \: \{1,N\} \]
Using Equation~\eqref{eq:3}, Equation~\eqref{eq:4}, Equation~\eqref{eq:5}, and Equation~\eqref{eq:7},  the total radiosity can be calculated as the contribution due to the NLOS radiosity and LOS radiosity as below, 
\[B_{total}=B_{NLOS}+B_{LOS}\]

Using these radiosity contributions, we can then solve the optimization problems formulated in the main paper, Section 4. The full steps are summarized in Algorithm~\ref{alg}.

%After the first bounce the radiosity is given by the condition  \eqref{eq:3} 
%\begin{equation} \label{eq:6}
%B_{nlos}= \rho_{nlos}B_iF_{ni}V_{nlos}  \quad \forall \:i\: \epsilon \: \{1,N\}  \quad 
%\end{equation}
%using  \eqref{eq:6} the radiosity at the end of the third bounce is  given by
%\begin{equation} \label{eq:7}
%B_{k}= \rho_{k}B_{nlos}F_{kn}V_{k}  \quad \forall \:i,k\: \epsilon \: \{1,N\}  \quad 
%\end{equation}
%Using \eqref{eq:3}, \eqref{eq:4}, \eqref{eq:5}, \eqref{eq:7},  the total radiosity is given as 
%\[B_{total}=B_{NLOS}+B_{LOS}\]

\begin{algorithm}
\SetAlgoLined
%\KwResult{Identify patch/patches to be illuminated along with corresponding illumination intensity per patch.}
 \textbf{Step 1:} Divide the LOS scene into $N$ patches, calculate the surface normal and area per patch. \newline
 \textbf{Step 2:} Calculate light source to LOS patch light transfer.
 \newline
  \For{LOS patch i=1:N}{
  Calculate  the first bounce radiosity: $B_i=\rho_iB_aF_{ia}$ where $B_a$ is the radiosity of the illumination source.
  }
  \textbf{Step 3:} LOS patch to the NLOS object light transfer. \newline
  \For{LOS patch i=1:N}{Calculate  the second NLOS bounce radiosity using Equation~\eqref{eq:4} and using Step 2 as the radiosity emitted by each LOS patch} 
  \textbf{Step 4:} Third bounce light from LOS to camera. \newline
  \For{LOS patch i=1:N}{Calculate  the final radiosity using Equation~\eqref{eq:7} } 
  \textbf{Step 5:} Solve corresponding optimization problem (Equation 2 or Equation 3 in the main paper) using radiosities from Step 4. 
 
 \caption{Adaptive Lighting to calculate $B_{NLOS}$}
 \label{alg}
\end{algorithm}

\begin{figure}[t]
\begin{center}
%\fbox{\rule{0pt}{2in} \rule{0.9\linewidth}{0pt}}
\includegraphics[width=0.5\columnwidth]{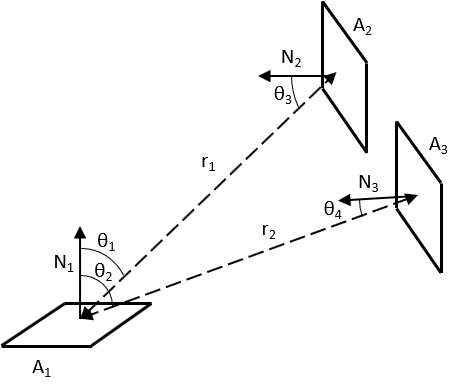}
\end{center}
   \caption{Radiosity measures the radiative transfer of light between diffuse surfaces and emitters based on their reflectance, viewing/occlusion, and geometric form-factors~\cite{cohen2012radiosity}. We formulate an optimization to identify the patches in the LOS which maximize the NLOS radiosity captured by the camera.  }
\label{fig:radiosity}
\end{figure}

% \begin{figure}[t]
% \begin{center}
% %\fbox{\rule{0pt}{2in} \rule{0.9\linewidth}{0pt}}
% \includegraphics[width=0.6\linewidth]{}IMAGES/voxel.png}
% \end{center}
%   \caption{We treat the problem of NLOS localization as a classification problem by dividing the occluded region into voxels and treating each voxel as a class, as visually depicted in this figure. }
% \label{fig:voxel}
% \end{figure}

\subsection{Energy-efficiency of Adaptive Lighting}

When a spatially-varying light pattern is used instead of a spotlight source, the illumination power is spread over the entire scene. This is counter to our stated goal of optimizing the energy-efficiency for the lighting, particularly the distributed optimization algorithm in the paper that operates under a power budget.  

\begin{figure}
\begin{center}
\includegraphics[width=0.5\linewidth]{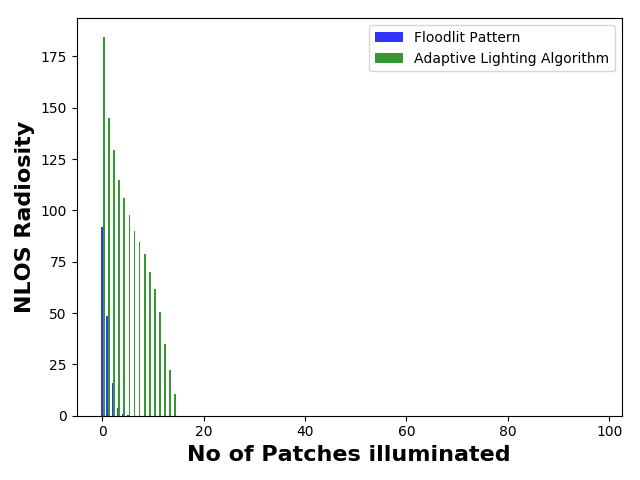}
\end{center}
   \caption{Plotting the NLOS radiosity returned by our adaptive lighting algorithm (green) versus floodlighting the scene (blue). Choosing optimal patches to illuminate returns more NLOS radiosity than spreading the illumination power evenly (as in a lighting pattern or floodlit) for the same illumination power.}
\label{fig:energyadaptive}
\end{figure}

To illustrate the effects of loss of power that occurs when you spread the light in a spatial pattern, we conducted the following experiment. Consider a room with the reflective wall divided into 100 patches. We illuminate the scene for a finite number of patches, and compare our adaptive lighting algorithm versus floodlighting the scene. For floodlighting, the incident illumination power is divided by the number of patches considered, and we measure the radiosity returned from NLOS. For our adaptive algorithm, we focus the incident illumination power onto a particular set of patches given by the optimization. In Figure~\ref{fig:energyadaptive}, we see that our adaptive algorithm (green) returns higher NLOS radiosity than spreading the light out in a floodlit pattern (blue). We believe this experiment shows the value of not using spatially-varying lighting patterns for the same energy budget. 

However, there is an interesting avenue for future research. One advantage is that spatially-varying lighting could improve detection coverage over the NLOS, as opposed to our adaptive lighting method which requires $N$ adaptive lighting patterns at test time to determine where the object is located. We can imagine NLOS imaging schemes which utilize spatially-varying lighting for coarse localization and detection, and then adaptive lighting for finer localization.

\section{Implementation Details}
For the simulated data, we implement our CNNs using PyTorch version 0.4.1 on a single NVIDIA GeForce GTX 1080 Ti GPU. Our datasets are of size 100,000 images for each specific wall and $64 \times 64$ resolution.  We train using stochastic gradient descent with momentum $0.9$ and learning rate $\lambda = 0.0001$ for $20$ epochs until convergence for classification and $16$ epochs until convergence for localization, with a $70:30$ training/testing split.

For our hardware prototype, we built a room setup, constructed using wood, of dimension 35.6 cm $\times$ 35.6 cm $\times$ 35.6 cm. We 3D printed the walls and then spray-painted them to be diffuse white. The real scene we use for the LOS is a variation of Wall 2. The Stanford bunny, sphere and man silhouette of varying sizes were 3D printed and spray painted diffuse white to help improve signal return back to the LOS. The wall was illuminated with an InFocus IN3138HD projector. We used an aperture after the projector of black construction paper with a small hole to focus the spot and emulate a spot light source. A Logitech C615 HD WebCam captured images of the diffuse wall. We capture roughly 10,000 real images to use for our datasets.

\subsection{Additional Real Data Localization Results}

 \begin{table}\footnotesize
        \begin{center}
        \begin{tabular}{|c|c|c|}%{ |M{20mm}|M{12mm}|M{9mm}|M{8mm}|M{8mm}| }
       
         \hline
         Object & \multicolumn{2}{c|}{Localization}  \\
      \cline{2-3}
    &Adaptive&Non-adaptive\\  
\cline{2-3}
      BUNNY 1  &2.41 cm &3.79 cm\\
      BUNNY 2 &1.32 cm &1.65 cm\\
      BALL 1&2.89 cm &4.67 cm\\
      BALL 2&1.61 cm&2.76 cm\\
    \hline
        \end{tabular}
           \end{center}
        \caption{ Additional real data results for four objects trained on the complex LOS wall 2.}
        \label{tab:table2}

    \end{table}

For our real data, we also performed an ablative study for localization with other sizes of spheres and bunnies. A bunny (BUNNY 1) with 5.5 cm width and 3.7 cm height was localized with MSE 2.41/3.79 cm respectively for adaptive/non-adaptive method, while a larger bunny (BUNNY 2) with 9 cm width and 7 cm height localized to 1.32/1.65 cm respectively for adaptive/non-adaptive method. A sphere (BALL 1) of diameter 3 cm localized to 2.89/4.67 cm respectively for adaptive/non-adaptive, and a larger sphere (BALL 2) of diameter 8 cm localized to 1.61/2.76 cm respectively for adaptive/non-adaptive. Note how as the size of objects gets bigger, the localization becomes more accurate in general due to more signal being reflected back from the NLOS.

\subsection{Saliency}

\begin{figure*}
\begin{center}
\includegraphics[width=0.8\linewidth]{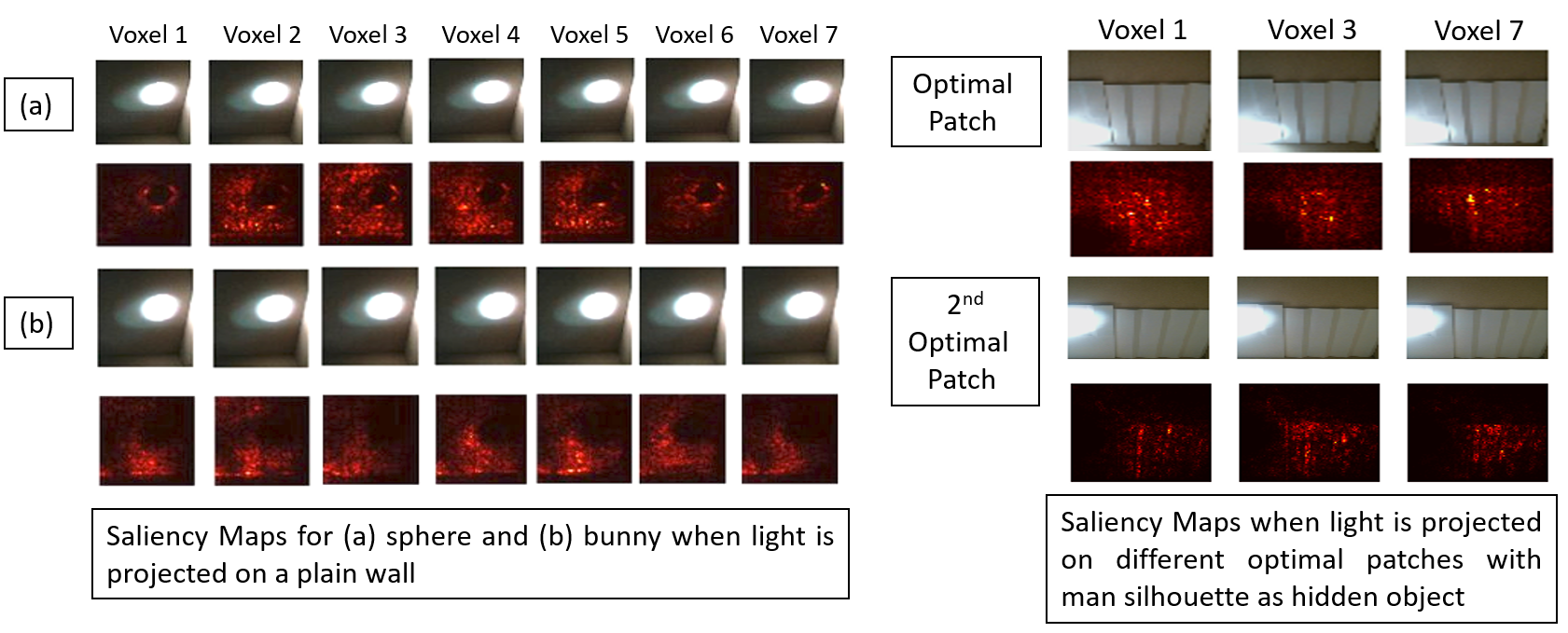}
\end{center}
   \caption{Salient image regions used by Inception network is calculated using the method from~\cite{simonyan2014very}.}
\label{fig:saliency}
\end{figure*}

To investigate what parts of the image our network is finding the most salient information, we utilize class-saliency maps from Simonyan et al.~\cite{simonyan2014very}. In Figure~\ref{fig:saliency}(a), we show the input images and saliency maps for the sphere and bunny projected on a planar wall. Note how the saliency of the sphere and bunny look qualitatively different, which probably explains why the network has poor generalisation performance across objects it has never seen in training before. In Figure~\ref{fig:saliency}(b), we show how the optimal patch returned by our adaptive lighting algorithm has more saliency for the network compared to the second best patch. This correlates with the improvement benefits we see with adaptive lighting.

\end{document}